\documentclass[aip,preprint]{revtex4-1}
\usepackage{graphicx}
\usepackage{dcolumn}
\usepackage{bm}

\begin{document}

\title{Controlled antivortex propagation at bifurcations in reconfigurable NdCo/NiFe racetracks} 

\author{V.V. Fernández}
\author{A.E. Herguedas-Alonso}
\affiliation{Depto. Física, Universidad de Oviedo, 33007 Oviedo, Spain.\\} 
\affiliation{CINN (CSIC–Universidad de Oviedo), El Entrego, Spain.\\}
\author{C. Fernandez-Gonzalez}
\author{R. Valcarcel}
\affiliation{ALBA Synchrotron, 08290 Cerdanyola del Vallès, Spain.\\}
\author{P. Suarez}
\author{A. G. Casero}
\affiliation{Depto. Física, Universidad de Oviedo, 33007 Oviedo, Spain.\\} 
\author{C. Quiros}
\affiliation{Depto. Física, Universidad de Oviedo, 33007 Oviedo, Spain.\\} 
\affiliation{CINN (CSIC–Universidad de Oviedo), El Entrego, Spain.\\}
\author{A. Sorrentino}
\affiliation{ALBA Synchrotron, 08290 Cerdanyola del Vallès, Spain.\\}
\author{A. Hierro-Rodríguez}
\email{hierroaurelio@uniovi.es}
\author{M. Vélez}
\email{mvelez@uniovi.es}
\affiliation{Depto. Física, Universidad de Oviedo, 33007 Oviedo, Spain.\\} 
\affiliation{CINN (CSIC–Universidad de Oviedo), El Entrego, Spain.\\}

\date{\today}

\begin{abstract}
The controlled propagation of spin textures at bifurcations is a critical challenge for racetrack-based logic devices. Here, we investigate the effect of longitudinal and transverse magnetic fields on the propagation of magnetic antivortices at bifurcations within the stripe domain pattern of a reconfigurable NdCo/NiFe racetrack in order to control the preferred antivortex trajectory. Magnetic Transmission X-ray Microscopy experiments were employed to correlate the observed propagation path with the local magnetic configuration. We demonstrate that Zeeman coupling to the magnetization components at the bifurcation core enables switching of the preferred propagation branch using low-amplitude transverse magnetic fields, without modifying the global stripe domain configuration that defines the guiding racetrack landscape. In-plane magnetic anisotropy provides an additional mechanism to break the symmetry between the upper and lower bifurcation branches by tuning the relative orientation between the stripe domain pattern and the longitudinal magnetic fields.

\end{abstract}

\pacs{}

\maketitle 
Magnetic racetracks, based on the controlled propagation of spin textures along predefined paths, are one of the most promising alternatives for the design of efficient memory and logic devices \cite{racetrack}.  In this context, bifurcations within the propagation path have been proposed as a basic logic unit for skyrmion \cite{skyrmion_gate} and domain wall (DW) racetracks \cite{DWlogic,DWlogic2,DWlogic3} with the successful demonstration of current driven DW logic gates in Y-shaped patterned wires \cite{Yjunction}  and probabilistic DW computing schemes in patterned nanowire networks \cite{Dedalo}.

A key question in each bifurcation type is the control of  path selectivity for spin texture propagation either deterministic \cite{biplexer} or probabilistic \cite{Dedalo2} depending on the relevant application. The texture chirality \cite{DWlogic3,Yjunction,biplexer}, the junction geometry \cite{DWlogic2}, and the topological transformations of spin textures at the junction \cite{Dedalo2,DWtopo} should be considered to understand the underlying mechanisms for path selection within the bifurcations. Additional factors \cite{DWtopo2} such as Walker breakdown or alternative magnetization reversal paths may also limit the available field and current ranges.

Recently, reconfigurable racetracks for vortex/antivortex (V/AV) textures have been demonstrated in NdCo$_{5}$/Ni$_{80}$Fe$_{20}$ bilayers\cite{PhysRevApplied23} based on two main ingredients: the first is rotatable anisotropy of stripe domains to fix reconfigurable propagation paths and the second is a very effective magnetic guiding mechanism for spin textures. Briefly, as sketched in Fig. 1(a), the NdCo$_{5}$ layer  defines the racetrack geometry with its periodic stripe domain pattern, controlled by in-plane saturating fields ($H_S$) and fixed at remanence ($H_R=0$) by rotatable anisotropy\cite{OriginRotAnis}. Then, spin textures propagating on the soft Ni$_{80}$Fe$_{20}$ layer are subject to a robust guiding mechanism\cite{Fernández_2026} based on a combination of magnetostatic interactions (between V/AV cores and the underlying $\pm m_z$ stripes) and topological restrictions (a lateral displacement into a neighboring stripe requires the absorption/emission of a Bloch point (BP) to switch spin texture polarity).

\begin{figure}
\includegraphics[width = 1\linewidth]{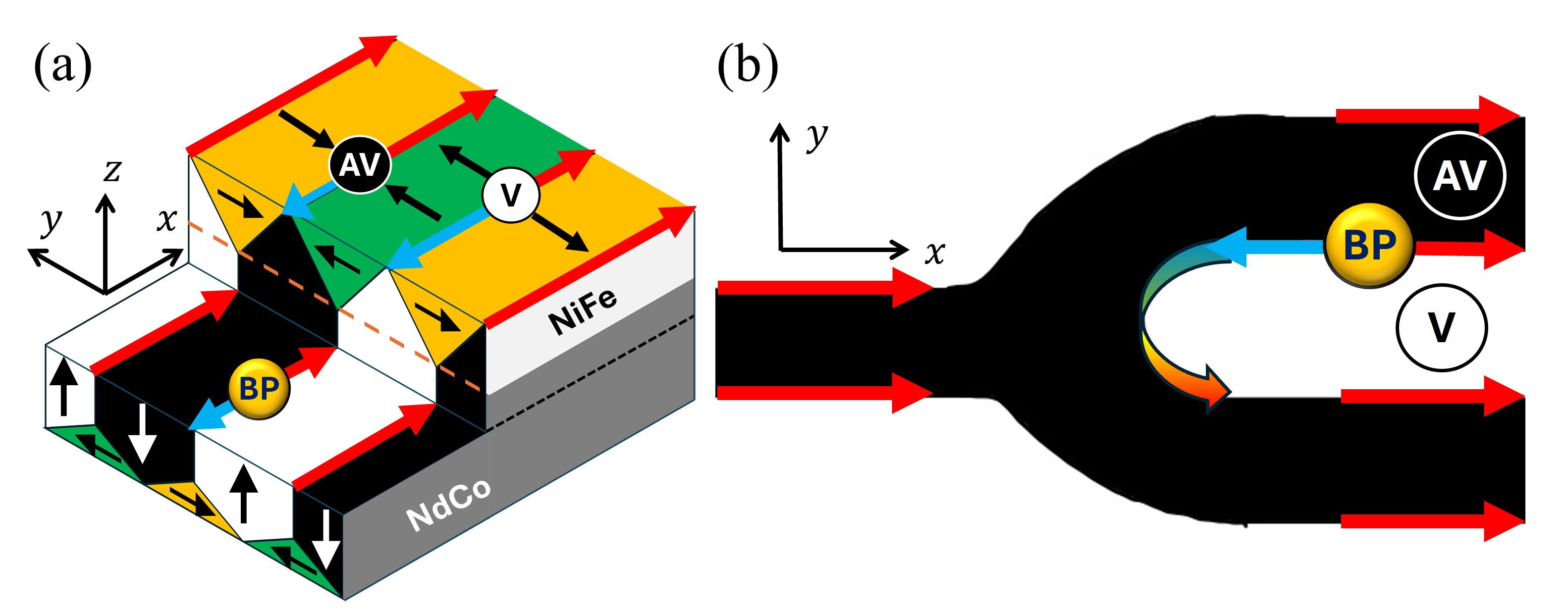}
\caption{\label{Fig1} (a) Sketch of a tail-to-tail DW within the stripe pattern of a NdCo/NiFe multilayer (arrows indicate the unit magnetization vector $\mathbf{m}=(m_x,m_y,m_z)$ at each domain):  at NdCo, there is a BP on the boundary between up/down stripes; at NiFe, the combination of $m_x$ reversal with $m_y$ signs given by the closure domain structure creates an AV with $-m_z$ core and a V with $+m_z$ core in adjacent lanes. (b) Sketch of a stripe bifurcation with an AV propagating along the upper bifurcation branch.}
\end{figure}

Bifurcations appear naturally in stripe patterns and are natural sites for spin texture nucleation during in-plane magnetization reversal  \cite{APL2017,PRB2017}. This process is subject to well defined topological constraints that rule the detailed configuration of Bloch lines, BPs, Vs and AVs\cite{Fernández_2026} that appear in the system. In particular, for a V/AV pair propagating away from a bifurcation core, the V always runs along the central stripe, whereas the AV moves along the bifurcated outer path\cite{APL2017} (see Fig.1(b)). 

Our goal here is to control the preferred path for AV propagation in order to design a toggle switch between the upper/lower bifurcation branches without altering the global stripe domain pattern that acts as the guiding landscape in the reconfigurable racetrack. For this purpose, external transverse fields can be a valuable tool to couple with magnetic textures within the bifurcation core: for example, it has been reported that in-plane magnetic fields can push Bloch lines within a Bloch DW\cite{PRB_BL,PRB_BLdyn} and, also, can tune dynamic topological transformation in stripe domain patterns\cite{stripesGDFE}. However, depending on the transverse field amplitude, they can also induce the motion of bifurcations within the sample\cite{Marisel} or even a global reorientation of the whole stripe pattern when the Zeeman coupling overcomes rotatable anisotropy\cite{Garnier_2020}.
 
In this work, spin texture propagation within stripe bifurcations of NdCo$_{5}$/Ni$_{80}$Fe$_{20}$ bilayers has been studied by Magnetic Transmission X-ray Microscopy (MTXM) under the effect of longitudinal ($H_x$) and transverse ($H_y$) fields. Clear correlations appear between the AV propagation branch and transverse field orientation, with the additional contribution of in-plane anisotropy to break $\pm y-$symmetry of the stripe pattern, which can be used to achieve robust control of AV propagation at stripe domain bifurcations. 

\begin{figure}
\includegraphics[width = 0.8\linewidth]{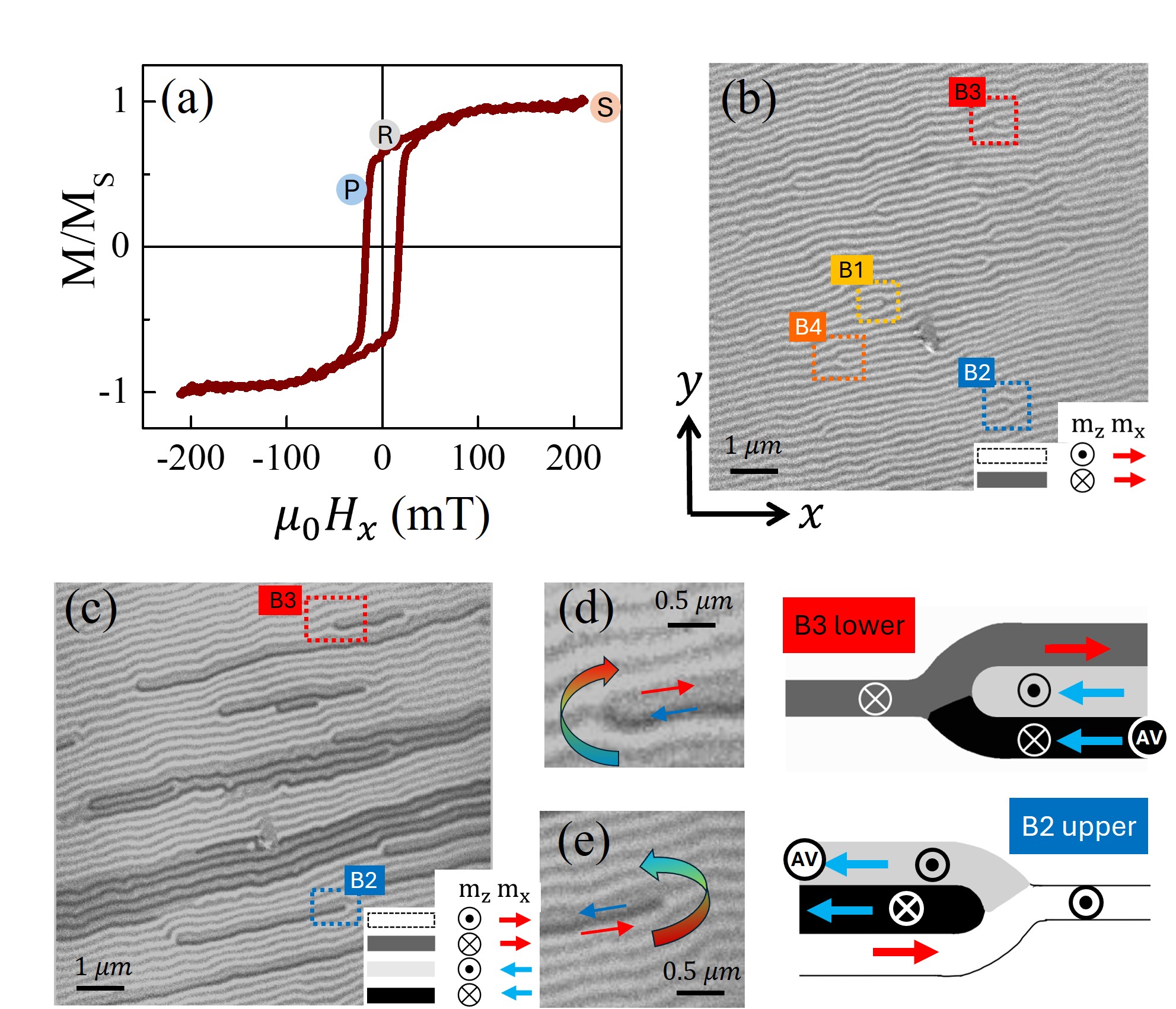}
\caption{\label{Fig2} (a) MOTKE hysteresis loop of a 80 nm NdCo$_5$/40 nm Ni$_{80}$Fe$_{20}$ bilayer. Labels indicate $H_x$ sequence used to nucleate and propagate spin textures: $H_S\rightarrow H_R\rightarrow H_P\rightarrow H_R)$.  (b-e) MTXM images at Fe L$_3$ edge with $\theta=30^\circ$, sensitive to $(m_x,m_z)$ within the Ni$_{80}$Fe$_{20}$ layer (contrast legend is indicated in the insets), and $H_y=0$: (b) initial remanence after $\mu_0H_S=+800$ mT, (c) partial reversal after $\mu_0 H_P= -13$ mT pulse. Dotted squares indicate different bifurcation types. (d-e) Zoom views of (c) showing B3-B2 bifurcations with partial in-plane magnetization reversal. Sketches show MTXM contrast depending on $(m_x,m_z)$ signs. Note that the circulation of magnetization around the core is directly related to the AV propagation path in each bifurcation. }
\end{figure}

80 nm NdCo$_{5}$/40 nm Ni$_{80}$Fe$_{20}$ bilayers  were grown  by dc magnetron sputtering\cite{PRB2017} on 50 nm thick $750\times 750$ $\mu$m$^2$ square Si$_3$N$_4$ membranes. NdCo$_5$ is an amorphous ferromagnetic alloy with saturation magnetization $M_S($NdCo$)=7.5\times 10^5$ A/m and weak perpendicular magnetic anisotropy $K_N=1.4 \times 10^5$ J/m$^3$ that creates the parallel stripe domain pattern at remanence \cite{huber}. The oblique deposition geometry of the sputtering chamber induces an in-plane anisotropy on NdCo$_5$ films\cite{jap_fvb,PhysRevApplied23} of the order of $K_u($NdCo$)=4-8\times10^3$ J/m$^3$ with its easy axis approximately $10^\circ$ away from the $x-$axis (defined by one of the sides of the square membrane). Permalloy Ni$_{80}$Fe$_{20}$ is a soft magnetic alloy with $M_S = 8.5 \times 10^5$ A/m and a small in-plane $K_u($NiFe$) = 850$ J/m$^3$. Figure 2(a) shows the hysteresis loop of the NdCo$_{5}$/Ni$_{80}$Fe$_{20}$ bilayer, measured by magnetooptical transverse Kerr effect (MOTKE). It displays the characteristic transcritical shape and reduced in-plane remanence associated with the $\pm m_z$ oscillation within the stripe domain pattern \cite{jap_fvb}.

For the MTXM experiments, the sample was taken to the full-field transmission microscope of the Mistral Beamline at Alba synchrotron \cite{NC2015} and illuminated with circularly polarized X-rays tuned to the Fe L$_3$ absorption edge at 706.8 eV to image the magnetic configuration at the NiFe layer. Magnetic contrast arises from Magnetic Circular Dichroism\cite{stöhr2023nature}, which is governed by the dot product of the spin angular momentum of the photons and the unit magnetization  vector ($\boldsymbol{\sigma}\cdot\boldsymbol{m}$). The sample was mounted with its $y$ axis perpendicular to the X-ray beam, so that $\boldsymbol{\sigma}\cdot\boldsymbol{m} \propto m_x \sin(\theta)+m_z \cos(\theta)$ with $\theta$ the angle of incidence of the X-ray beam. Thus, MTXM images contain information on both in-plane $m_x$ and out-of-plane $m_z$ magnetic configuration. The sample holder counts with a pair of coils to generate longitudinal pulsed fields up to $\mu_0 H_x= 2000$ mT (with 16 $\mu$s pulse width). Variable transverse DC fields up to $\mu_0 H_y = 5$ mT were also applied \textit{in situ} with the aid of an adjustable permanent magnet. 

The propagation of spin textures at stripe domain bifurcations has been studied following a two-step field sequence. First, the stripe orientation is configured with a large pulsed field $H_S$ along the $x$-axis. Figure 2(b) shows the MTXM in this initial remanent state ($H_R=0$), acquired with oblique incidence ($\theta=30^\circ$) after a $\mu_0H_S=+800$ mT pulse. It presents a clear pattern of stripes with only two different contrast levels, indicating a magnetic configuration within the NiFe layer with uniform $m_x$ sign ($+m_x$ as set by $H_S$) combined with a weak $\pm m_z$ oscillation, imprinted by the stripe domains in the NdCo layer. Note the presence of different stripe bifurcations in the sample (see the dotted rectangles in Fig. 2(b)), which can be classified into four types (B1-B4) depending on whether the Y-shaped stripe opens to the right or to the left and its $m_z$ sign. 

\begin{table*}
\begin{ruledtabular}
    \centering
    \caption{Bifurcation core statistics as a function of $\mu_0H_S$ and $\mu_0H_y$ within 24 $10\times 10 \mu$m$^2$ regions of the NdCo/NiFe bilayer.}
    \label{tab:placeholder_label}
    \begin{tabular}{ccccccc}   
        $\mu_0 H_{S}/\mu_0H_P$ (mT)& bifurcation& AV branch& $m_y$ @ core& $\mu_0H_y =-5$ mT& $\mu_0H_y=0$ & $\mu_0H_y=+5$ mT \\    \hline  
        +800/-13& $B_2$ & upper& $+$& 0& 1& 12\\  
        +800/-13& $B_2$ & lower& $-$& 5& 0& 0\\ \hline
        +800/-13& $B_3$ & upper & $-$& 11& 0& 0\\    
        +800/-13& $B_3$& lower& $+$& 0& 6& 7\\ \hline
        -800/+13& $B_1$& upper & $-$& 8& 3& 0\\
        -800/+13& $B_1$& lower& $+$& 0& 0& 21\\ \hline
        -800/+13& $B_4$& upper & $+$& 0& 0& 13\\      
        -800/+13& $B_4$& lower& $-$& 32& 14& 0\\     
    \end{tabular}
\end{ruledtabular}
\end{table*}

In a second step, a DC $H_y$ field is applied, followed by a longitudinal field pulse $H_P$ (with opposite polarity to $H_S$) to induce the propagation of Vs and AVs away from bifurcation cores. Indeed, the MTXM image in Fig. 2(c), acquired at remanence after a $\mu_0H_P=-13$ mT pulse with $H_y=0$, shows four different contrast levels corresponding to the four possible  combinations of $(\pm m_x,\pm m_z)$. The detailed analysis of MTXM contrasts close to partially reversed bifurcations (see Figs. 2(d-e) for bifurcations B2 and B3) allows us to correlate the AV propagation path with the magnetic configuration at each bifurcation core, as indicated in the accompanying sketches. For example, at the B3 bifurcation shown in Fig. 2 (d), we can observe a black contrast stripe (corresponding to $(-m_x,-m_z)$). It indicates that an AV has moved away from B3 along the lower bifurcation branch, which, by continuity, implies a clock-wise rotation of the magnetization around the core (see the curved arrow in Fig. 2(d)). Thus, it corresponds to a $+m_y$ orientation at the core. A similar analysis of the B2 bifurcation in Fig. 2(e) indicates that the AV has propagated along the upper bifurcation branch leaving behind a $+m_y$ orientation at the core.

Table 1 is a summary of AV propagation statistics as a function of $H_S$ and  $H_y$ signs, measured in 24 different $10\times10$ $\mu$m$^2$ sample regions. Depending on $H_S$ sign, $m_x$ reversal within the NiFe layer can occur only at two of the four possible bifurcation types \cite{PRB2017} (B2/B3 for $H_S>0$ and B1/B4 for $H_S<0$). In each case, the  orientation of $m_y$ at the core after AV propagation has been determined by continuity, as sketched in Fig. 2. Then, the number of partially reversed bifurcations of each possible configuration has been recorded for $\mu_0H_S=\pm 800$ mT and $\mu_0H_y=0$ and $\pm5$ mT. A clear correlation appears between the orientation of $m_y$ at the bifurcation core and the sign of $H_y$: at $\mu_0H_y= +5$ mT, $m_y$ is positive at 100\% of the recorded bifurcation cores independent of $H_S$ (while negative $m_y$ cores are always observed at $\mu_0H_y=-5$ mT). This suggests that Zeeman coupling between the transverse field and the magnetization at the bifurcation core has a direct influence on the preferred branch for AV propagation in each given bifurcation. For $H_y=0$, the sign of $H_S$ appears as the relevant factor to determine $m_y$ at the bifurcation core (with positive $m_y$ observed for positive $H_S$ and vice versa in table 1).

\begin{figure}
\includegraphics[width = 0.8\linewidth]{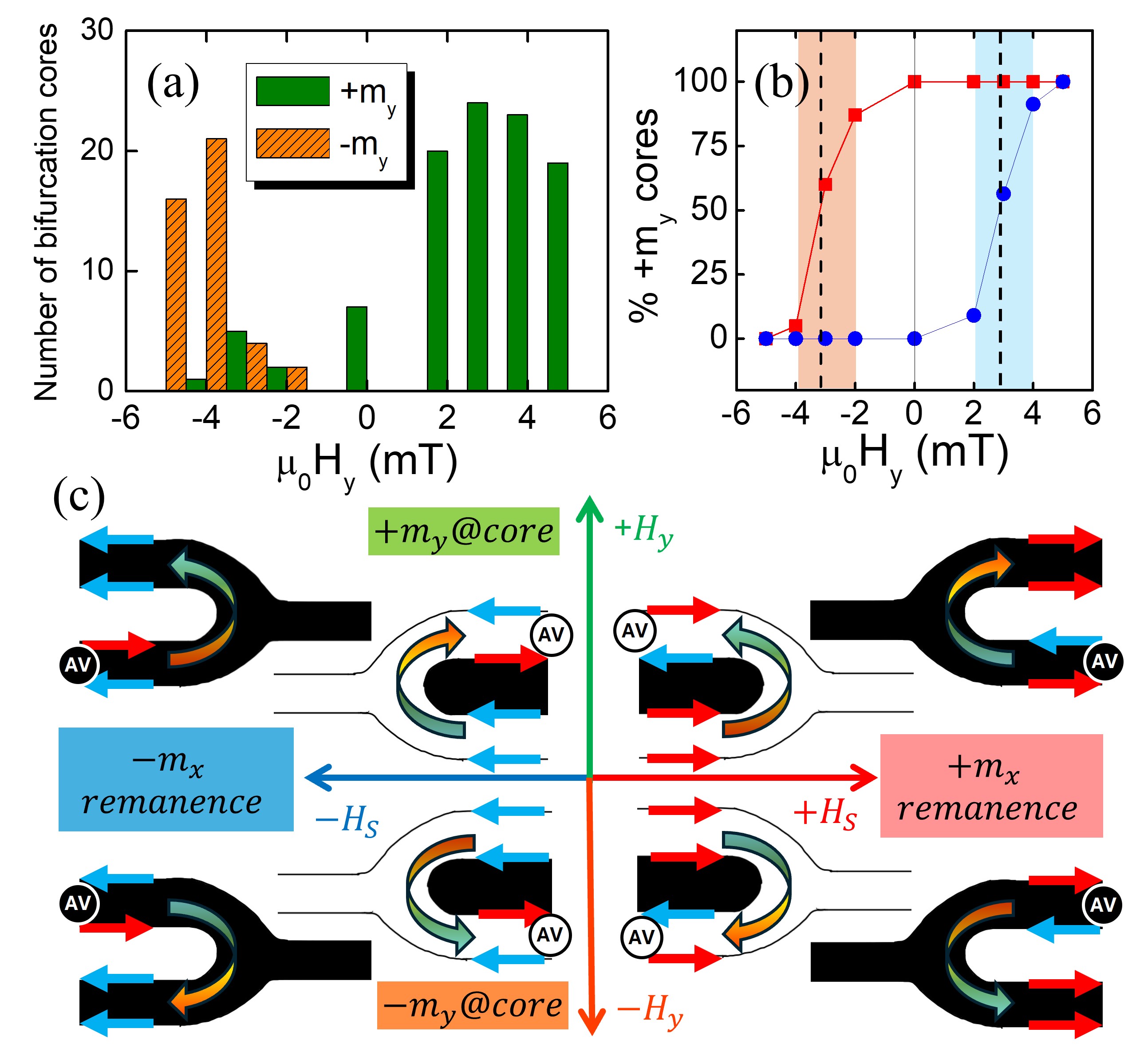}
\caption{\label{Fig3} (a) Statistics of $m_y$ orientation at bifurcation cores as a function of transverse dc field with $\mu_0H_S=+800$ mT and $\mu_0 H_P=- 13$ mT; (b) Fraction of $+m_y$ bifurcation cores vs. $\mu_0 H_y$: red squares ($\mu_0 H_S = +800$ mT, $\mu_0 H_P=- 13$ mT), blue dots ($\mu_0H_S =-800$ mT, $\mu_0 H_P=+ 13$ mT). Note that bifurcations with AV propagation beyond the field of view have also been included to increase the number of events. Dashed lines indicates $\mu_0 H_y^{*}$, corresponding to 50\% $\pm m_y$ probability; shaded areas indicate the field intervals $\Delta\mu_0H_y$ with mixed presence of $\pm m_y$ cores for each $H_S$ polarity. (c) $H_y$ \textit{vs.} $H_S$ diagram of preferred bifurcation branches for AV propagation. }
\end{figure}

The evolution in the number of $\pm m_y$ bifurcation cores \textit{vs.} $\mu_0 H_y$, shown in Fig. 3(a) for $\mu_0H_S=+800$ mT, reveals three different regimes: in the range $5$ mT $\geq \mu_0H_y\geq0$, $m_y$ is positive in all partially reversed bifurcations;  for $-2$ mT $\geq \mu_0H_y\geq-4$ mT, there is a variable mixture of bifurcations with $\pm m_y$ orientation at the core; and, finally, at $\mu_0H_y=-5$ mT, $m_y$ is negative in all cases. This gradual transition from $+m_y$ to $-m_y$ bifurcation cores as a function of $\mu_0H_y$ appears for both $H_S$ orientations (see Fig. 3(b)): centered in a finite $|\mu_0H_y^*|\approx 3$ mT, corresponding to a 50\% probability of $\pm m_y$ cores, and spanning over a similar transverse field interval ($\Delta\mu_0H_y\approx 2$ mT). Thus, transverse field amplitudes above 5 mT ensure a strong enough Zeeman coupling with the magnetization at the bifurcation core to provide a robust control on the preferred branches for AV propagation in each bifurcation type (as summarized in the diagram in Fig. 3(c) for the different $H_S$, $H_y$ combinations).
 
\begin{figure}
\includegraphics[width = 0.7\linewidth]{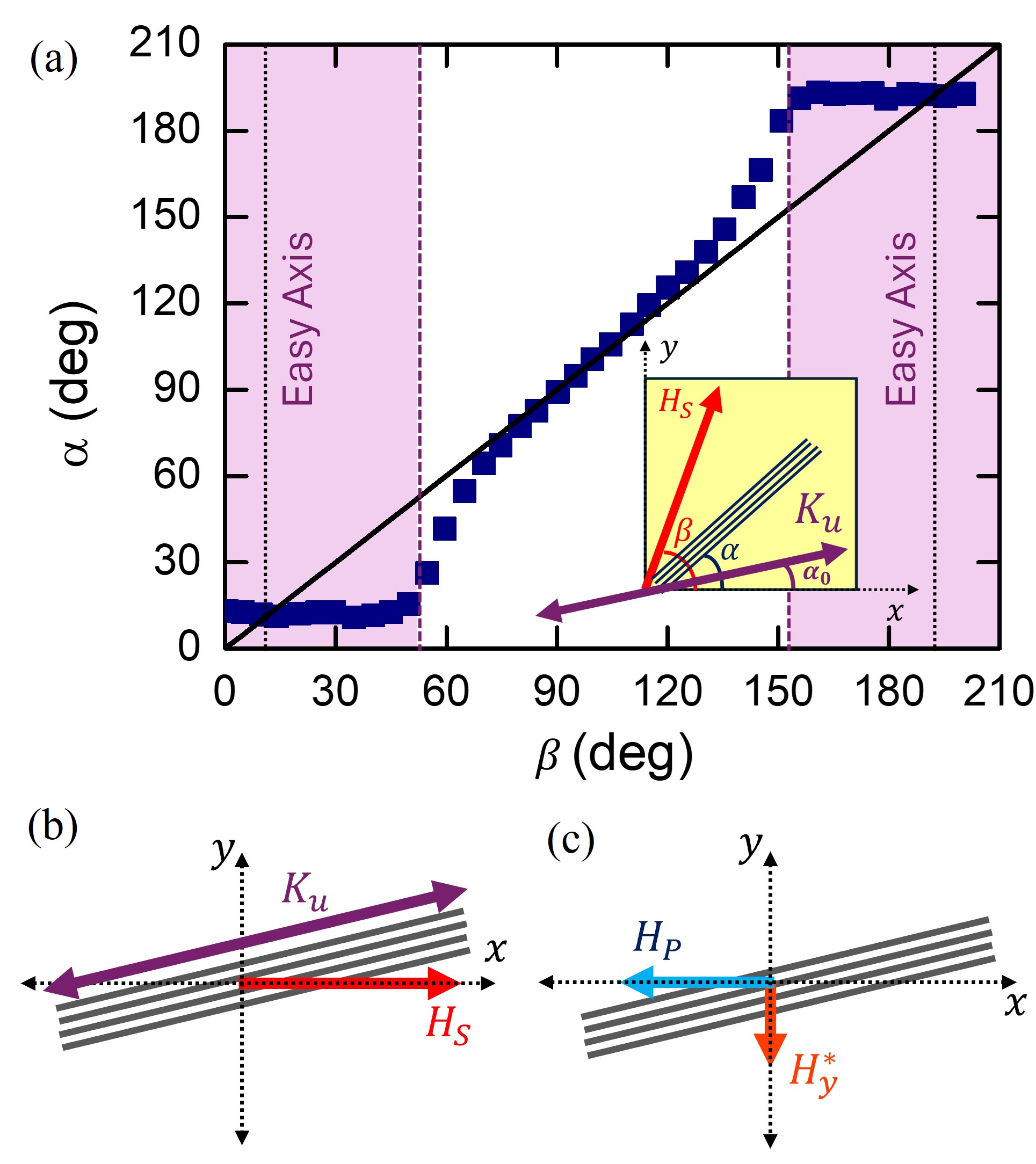}
\caption{\label{Fig4} (a) Orientation  of stripe domains at remanence ($\alpha$) \textit{vs.} $H_S$ field orientation ($\beta$) with $\mu_0 H_S= 150$ mT. Inset is a sketch of the magnetic field and stripe domain geometry relative to the membrane frame and to the easy axis of $K_u($NdCo$)$. Solid line indicates $\alpha=\beta$, corresponding to stripe pattern orientation under pure rotatable anisotropy. Shaded area indicates the angular region dominated by $K_u($NdCo$)$. (b-c) Sketches of orientation of the stripe pattern at remanence relative to in-plane anisotropy axis at $\alpha_0=12^\circ$ and the fields applied during the AV propagation experiment ($H_S$ and $H_P$ at $\beta=0^\circ$ and $H_y$ at $\beta=90^\circ$).} 
\end{figure}

The data in Table 1 and Fig. 3 show the presence of an additional mechanism that breaks $\pm y$ symmetry at the bifurcations in the absence of transverse fields, favoring a specific $m_y$ orientation linked to the longitudinal $H_S$ field. This can be traced back to the interplay between $H_S$ and in-plane uniaxial anisotropy $K_u($NdCo$)$. Figure 4(a) shows  the orientation of stripe domains at remanence ($\alpha$) as a function of the $H_S$ field orientation ($\beta$) obtained by Kerr effect microscopy (both angles are measured relative to the $x-$axis defined by the square frame of the membrane, as indicated in the inset of Fig. 4(a)). When $H_S$ is applied near the $x-$axis ($\beta$ close to $0^\circ$ and $180^\circ$), the orientation of the stripes remains fixed at $\alpha_0=12^\circ$, indicating the prevalence of $K_u($NdCo$)$ to turn the average in-plane magnetization of the stripe pattern towards its easy axis\cite{jap_fvb,PRB2013}. At higher $\beta$, the effect of $K_u($NdCo$)$ weakens and the stripe pattern starts to follow the applied field orientation, approaching the $\alpha=\beta$ line characteristic of pure rotatable anisotropy (i.e. easy axis parallel to $H_S$ \cite{OriginRotAnis}).   

Therefore, in the low angle region dominated by $K_u($NdCo$)$ (shaded area in Fig. 4(a)), in-plane anisotropy can be used as a static mechanism to set the preference for either $+m_y$ or $-m_y$ configuration at bifurcation cores, playing only with the angle between $H_S$ and the easy axis $\beta-\alpha_0$. In our case, with $H_S$ at $\beta=0$ and the easy axis at $\alpha_0=12^\circ$ (see sketches in Figs. 4(b-c)), the fact that the stripe pattern is aligned with the $K_u($NdCo$)$ easy axis (i.e. $\alpha=\alpha_0$) implies that the reversed field $\mu_0H_P=-13$ mT has a positive transverse component relative to the stripe pattern ($\mu_0H_\perp=\mu_0H_P \sin(\beta-\alpha_0)=2.1$ mT) that favors $+m_y$ bifurcation cores, consistent with the data in Table 1. Then, a finite negative $\mu_0H_y$ is needed to achieve a symmetric configuration at the bifurcations with the magnetic field fully aligned with the stripe pattern. In particular, the experimental $\mu_0 H_y^{*}\approx -3$ mT, needed for equal probability of $\pm m_y$ cores in the data of Fig. 3(b), corresponds to an effective total field $\mu_0 \mathbf{H} =(\mu_0 H_P, \mu_0 H_y^{*}) = (-13,-3)$ mT oriented at $\beta=13^\circ$, i.e. along the $K_u($NdCo$)$ easy axis. 

In summary, the propagation of spin textures at bifurcations within a stripe domain pattern has been studied by MTXM experiments in a  NdCo/NiFe bilayer. It is found that the preferred propagation path for AVs at each bifurcation kind can be tuned through Zeeman coupling to $m_y$ magnetization components at the bifurcation core by two different mechanisms. First, direct coupling to low amplitude transverse  $\mu_0H_y$ fields can toggle the $\pm m_y$ preference at bifurcation cores without altering the overall stripe domain pattern. Second, in-plane anisotropy  $K_u($NdCo$)$ can be used to tune the relative orientation between $H_S$ and the stripe pattern at remanence ($\beta-\alpha$), resulting in effective transverse field components that break $\pm y-$symmetry at bifurcations. Both mechanisms provide full control of AV propagation branch at bifurcations (100\% probability) except for a small transverse field range ($\Delta\mu_0H_y\approx 2$ mT) centered at the ideal symmetric configuration with the reverse field parallel to the stripe domain pattern (i.e. $\mu_0 \mathbf{H} =(\mu_0 H_P, \mu_0 H_y^{*})$). Therefore, these results demonstrate deterministic control over spin-texture propagation at stripe-domain bifurcations, highlighting their potential as reconfigurable logic units within NdCo/NiFe racetracks.

\begin{acknowledgments}
VVF acknowledges support from the Severo Ochoa Predoctoral Fellowship Program (PA-22-BP21-124). This work has been supported by Spanish MCIN/AEI/ 10.13039/ 501100011033/ FEDER, UE under grant PID2022-136784NB and by Agencia SEKUENS (Asturias) under grant UONANO IDE/2024/000678 with the support of FEDER funds.
\end{acknowledgments}

\section*{Data Availability Statement}

The data that support the findings of this study are openly available in RUO (Repositorio Institucional de la Universidad de Oviedo) at https://digibuo.uniovi.es/
 
\bibliography{CWmeron}

@article{racetrack,
     author={Bläsing, Robin and Khan, Asif Ali and Filippou, Panagiotis Ch. and Garg, Chirag and Hameed, Fazal and Castrillon, Jeronimo and Parkin, Stuart S. P.},
  journal={Proceedings of the IEEE}, 
  title={Magnetic Racetrack Memory: From Physics to the Cusp of Applications Within a Decade}, 
  year={2020},
  volume={108},
  number={8},
  pages={1303-1321}}

@article{skyrmion_gate,
author = {Zhang, X. and Ezawa, M. and Zhou, Y.}, 
title = {Magnetic skyrmion logic gates: conversion, duplication and merging of skyrmions},
journal = {Scientific Reports},
volume = {5},
pages = {9400},
year = {2015}}

@article{DWlogic,
author = {Luo, Z. and Hrabec, A. and Dao, T.P. and G. Sala and 
S. Finizio and J. Feng and S. Mayr and J. Raabe and P. Gambardella and L.J. Heyderman }, 
title = {Current-driven magnetic domain-wall logic},
journal = {Nature},
volume = {579},
pages = {214–218},
year = {2020}}

@article{biplexer,
    author = {Phung, Timothy and Pushp, Aakash and Rettner, Charles and Hughes, Brian P. and Yang, See-Hun and Parkin, Stuart S. P.},
    title = {Robust sorting of chiral domain walls in a racetrack biplexer},
    journal = {Applied Physics Letters},
    volume = {105},
    pages = {222404},
    year = {2014}    
}

@article{Yjunction,
author = {Garg, Chirag and Pushp, Aakash and Yang, See-Hun and Phung, Timothy and Hughes, Brian P. and Rettner, Charles and Parkin, Stuart S. P.},
title = {Highly Asymmetric Chiral Domain-Wall Velocities in Y-Shaped Junctions},
journal = {Nano Letters},
volume = {18},
number = {3},
pages = {1826-1830},
year = {2018}}

@article{DWlogic2,
author = {Şteţco, Elena M. and Petrişor, Traian Jr. and Pop, Ovidiu A. and Belmeguenai, Mohamed and Miron, Ioan M. and Gabor, Mihai S.},
title = {Diode and Selective Routing Functionalities Controlled by Geometry in Current-Induced Spin–Orbit Torque Driven Magnetic Domain Wall Devices},
journal = {Nano Letters},
volume = {24},
number = {44},
pages = {13991-13997},
year = {2024}}

@article{Dedalo,
author = {Sanz-Hernández, Dédalo and Massouras, Maryam and Reyren, Nicolas and Rougemaille, Nicolas and Schánilec, Vojtěch and Bouzehouane, Karim and Hehn, Michel and Canals, Benjamin and Querlioz, Damien and Grollier, Julie and Montaigne, François and Lacour, Daniel},
title = {Tunable Stochasticity in an Artificial Spin Network},
journal = {Advanced Materials},
volume = {33},
number = {17},
pages = {2008135},
year = {2021}
}

@article{Dedalo2,
  title = {Real Space Imaging of Field-Driven Decision-Making in Nanomagnetic {G}alton Boards},
  author = {Arava, H. and Sanz-Hernandez, D. and Grollier, J. and Petford-Long, A. K.},
  journal = {Phys. Rev. Lett.},
  volume = {134},
  issue = {8},
  pages = {086704},
  numpages = {6},
  year = {2025},
  }

@article{DWlogic3,
author = {Omari, Khalid A. and Broomhall, Thomas J. and Dawidek, Richard W. S. and Allwood, Dan A. and Bradley, Ruth C. and Wood, Jonathan M. and Fry, Paul W. and Rosamond, Mark C. and Linfield, Edmund H. and Im, Mi-Young and Fischer, Peter J. and Hayward, Tom J.},
title = {Toward Chirality-Encoded Domain Wall Logic},
journal = {Advanced Functional Materials},
volume = {29},
pages = {1807282},
year = {2019}}

@article{DWtopo,
author = {Pushp, A. and Phung, T. and Rettner, C. and B. P. Hughes and S.-H. Yang and L. Thomas and S. S. P. Parkin  },
title = {Domain wall trajectory determined by its fractional topological edge defects},
journal = {Nature Physics},
volume = {9},
pages = {505–511},
year = {2013}}

@article{DWtopo2,
author = {S K Walton and K Zeissler and D M Burn and S Ladak and D E Read and T Tyliszczak and L F Cohen and W R Branford },
title = {Limitations in artificial spin ice path selectivity: the challenges beyond topological control},
journal = {New Journal of Physics},
volume = {17},
pages = {013054},
year = {2015}}

@article{Fernández_2026,
year = {2025},
volume = {9},
pages = {015002},
author = {Fernández, V V and Ferrer, S and Hierro-Rodríguez, A and Vélez, M},
title = {Nucleation of magnetic textures in stripe domain bifurcations for reconfigurable domain wall racetracks},
journal = {Journal of Physics: Materials}
}

@article{Garnier_2020,
year = {2020},
volume = {3},
pages = {024001},
author = {Garnier, Louis-Charles and Marangolo, Massimiliano and Eddrief, Mahmoud and Bisero, Diego and Fin, Samuele and Casoli, Francesca and Pini, Maria Gloria and Rettori, Angelo and Tacchi, Silvia},
title = {Stripe domains reorientation in ferromagnetic films with perpendicular magnetic anisotropy},
journal = {Journal of Physics: Materials}}

@article{PRB_BL,
  title = {Domain walls within domain walls in wide ferromagnetic strips},
  author = {Herranen, Touko and Laurson, Lasse},
  journal = {Phys. Rev. B},
  volume = {92},
  pages = {100405},
  year = {2015},
  }

@article{PRB_BLdyn,
  title = {Bloch-line dynamics within moving domain walls in 3D ferromagnets},
  author = {Herranen, Touko and Laurson, Lasse},
  journal = {Phys. Rev. B},
  volume = {96},
  pages = {144422},
  year = {2017},
  }

@article{stripesGDFE,
author = {Titze, Tim and Koraltan, Sabri and Schmidt, Timo and Matthies, Mailin and Fernández-Pacheco, Amalio and Suess, Dieter and Albrecht, Manfred and Mathias, Stefan and Steil, Daniel},
title = {Controlling Bubble and Skyrmion Lattice Order and Dynamics via Stripe Domain Engineering in Ferrimagnetic Fe/Gd Multilayers},
journal = {Advanced Physics Research},
year = {2025},
pages = {e00194}}

@article{Marisel,
  title = {Unidirectional motion of topological defects mediating continuous rotation processes},
author = {M. Di Pietro Martínez and L. A. Turnbull and J. Neethirajan and M. Birch and S. Finizio and J. Raabe and E. Lesne and A. Markou and M. Vélez and A. Hierro-Rodríguez and M, Salvalaglio and C. Donnelly},
  journal = {npj Spintronics},
volume = {3},
pages = {47},
year = {2025},
  }

@article{PhysRevApplied23,
  title = {Memory effects on the current-induced propagation of spin textures in {N}d{C}${\mathrm{o}}_{5}$/{N}$\mathrm{i}_{8}${F}$\mathrm{e}_{2}$ bilayers},
  author = {Fern\'andez, V.V. and Herguedas-Alonso, A.E. and Hermosa, J. and Aballe, L. and Sorrentino, A. and Valcarcel, R. and Quiros, C. and Mart\'{\i}n, J.I. and Pereiro, E. and Ferrer, S. and Hierro-Rodr\'{\i}guez, A. and V\'elez, M.},
  journal = {Phys. Rev. Appl.},
  volume = {23},
  issue = {1},
  pages = {014023},
  numpages = {14},
  year = {2025},
  
}

@article{PRB2017, 
title = {Observation of asymmetric distributions of magnetic singularities across magnetic multilayers},
  author = {Hierro-Rodriguez, A. and Quir\'os, C. and Sorrentino, A. and Blanco-Rold\'an, C. and Alvarez-Prado, L. M. and Mart\'{\i}n, J. I. and Alameda, J. M. and Pereiro, E. and V\'elez, M. and Ferrer, S.},
  journal = {Phys. Rev. B},
  volume = {95},
  issue = {1},
  pages = {014430},
  numpages = {9},
  year = {2017},
 }

@book{huber, 
title = {Magnetic Domains: The Analysis of Magnetic Microstructures},
  author={A. Hubert and R. Schäfer},
      year={1998},
  publisher={Springer-Verlag, Berlin}}

@article{jap_fvb,
    author = {Valdés-Bango, F. and García Alonso, F. J. and Rodríguez-Rodríguez, G. and Morán Fernandez, L. and Anillo, A. and Ruiz-Valdepeñas, L. and Navarro, E. and Vicent, J. L. and Vélez, M. and Martín, J. I. and Alameda, J. M.},
    title = "{Perpendicular magnetic anisotropy in Nd-Co alloy films nanostructured by di-block copolymer templates}",
    journal = {Journal of Applied Physics},
    volume = {112},
       pages = {083914},
    year = {2012},
    }

@article{OriginRotAnis,
    author = {Fujiwara, H. and Sugita, Y. and Saito, N.},
    title = "{Mechanism of Rotatable Anisotropy in Thin Magnetic Films of {N}i, {F}e, and {N}i–{F}e}",
    journal = {Applied Physics Letters},
    volume = {4},
    number = {12},
    pages = {199-200},
    year = {1964},
    month = {06},
    issn = {0003-6951},
    doi = {10.1063/1.1753938},
}

@article{NC2015, 
author={C. Blanco-Roldán and C. Quirós and A. Sorrentino and A. Hierro-Rodríguez and L. M. Álvarez-Prado and R. Valcárcel and M. Duch and N. Torras and J. Esteve and J. I. Martín and M. Vélez and J. M. Alameda and E. Pereiro and S. Ferrer},
title = "{Nanoscale imaging of buried topological defects with quantitative X-ray magnetic microscopy}",
journal = {Nat. Commun.},
    volume = {6},
        pages = {8196},
    year = {2015}}

@book{stöhr2023nature,
  title={The Nature of X-Rays and Their Interactions with Matter},
  author={St{\"o}hr, J.},
  isbn={9783031207440},
  series={Springer Tracts in Modern Physics},
  year={2023},
  publisher={Springer International Publishing}}

@article{APL2017,
    author = {Hierro-Rodriguez, A. and Quirós, C. and Sorrentino, A. and Valcárcel, R. and Estébanez, I. and Alvarez-Prado, L. M. and Martín, J. I. and Alameda, J. M. and Pereiro, E. and Vélez, M. and Ferrer, S.},
    title = "{Deterministic propagation of vortex-antivortex pairs in magnetic trilayers}",
    journal = {Applied Physics Letters},
    volume = {110},
    number = {26},
    pages = {262402},
    year = {2017},
   
}

@article{PRB2013,
  title = {Controlled nucleation of topological defects in the stripe domain patterns of lateral multilayers with perpendicular magnetic anisotropy},
  author = {Hierro-Rodriguez, A. and V\'elez, M. and Morales, R. and Soriano, N. and Rodr\'{\i}guez-Rodr\'{\i}guez, G. and \'Alvarez-Prado, L. M. and Mart\'{\i}n, J. I. and Alameda, J. M.},
  journal = {Phys. Rev. B},
  volume = {88},
  pages = {174411},
    year = {2013},
}

\end{document}